\documentclass[singlecolumn,11pt]{llncs}
\usepackage{graphicx}
\usepackage{dcolumn}
\usepackage{rotating}
\usepackage{supertabular}
\usepackage{longtable}
\usepackage{bm}
\usepackage{verbatim}
\usepackage{textcomp}
\usepackage{amsfonts}
\usepackage{listings}

\begin{document}
\title{Interference Automata}
\author{M. V. Panduranga Rao\thanks{A preliminary version of this paper appears in the proceedings
of the 16th Australasian Workshop on Combinatorial Algorithms (AWOCA 2005).}}
\institute{Department of Computer Science and Automation \\Indian Institute of Science\\ Bangalore  560 012\\ India  \\\email{pandurang@csa.iisc.ernet.in}}
\date{\today}

\maketitle
\begin{abstract}
We propose a computing model, the Two-Way Optical Interference Automata (2OIA), 
that makes use of the phenomenon of optical interference.
We introduce this model to investigate the increase in power, in terms of language recognition, of a classical
Deterministic Finite Automaton (DFA) when endowed with the facility of optical interference.
The question is in the spirit of Two-Way Finite Automata With Quantum and Classical States (2QCFA)
[A. Ambainis and J. Watrous, Two-way Finite Automata With Quantum and Classical States, Theoretical Computer Science, 
287 (1), 299-311, (2002)] wherein the classical DFA
is augmented with a quantum component of constant size. 
We test the power of 2OIA against the languages mentioned in the above paper.
We give efficient 2OIA algorithms to recognize languages for which 
2QCFA machines have been shown to exist, as well as languages whose status vis-a-vis 2QCFA
has been posed as open questions. Finally we show the existence of a language that
cannot be recognized by a 2OIA but can be recognized by an $O(n^3)$ space Turing machine.

\end{abstract}

We present a model of automata, the \emph{Two-Way Optical Interference Automata} (OIA), that uses the phenomenon of interference 
to recognize languages.

We augment the classical 2DFA with an array of sources of 
monochromatic light and a detector. 
The guiding principle behind the design of the 2OIA model is to deny it any
resource other than a finite control and the ability to create interference.
Our interest lies essentially in wave interference; for concreteness and ease of
exposition we choose light.

Specifically, we address the following question: given a language $L\subseteq \{a,b\}^*$, an input $w\in \{a,b\}^*$,
and a 2DFA augmented with $|w|$ ``sources of interference", is it possible to decide efficiently if $w\in L$ by examining their interference patterns? 
While this question is interesting in its own right, the model abstracts out the phenomenon
of interference from quantum automata models~\cite{MCr,KW,BP,AK,Dense,AF} in the most general sense.

A typical example of an automata model having a restricted quantum component is the 
2-way Finite Automata model with Quantum and Classical states (2QCFA) of Ambainis and 
Watrous~\cite{AW}, which is essentially a classical 2DFA that reads input off a 
read-only tape and is augmented with a quantum component of
constant size: the number of dimensions of the associated Hilbert space does not
depend on the input length. Unitary operations on the quantum component, which are
performed interleaved with classical transitions, evolve the quantum state vector,
producing interference in probability amplitudes of the vector. The classical
component takes the result of measurement operations on the quantum state vector
into account while deciding membership of a given input string.
Ambainis and Watrous \cite{AW} showed 2QCFA that accept $L_{eq}=\{a^nb^n \mid n \in \mathbb{N}\}$ in polynomial time and
palindromes ($L_{pal}=\{ww^R\mid w\in \{a,b\}^*$ where $w^R$ is $w$ reversed$\}$)
in exponential time with bounded error.

The interference produced by 
the sources in our model is the analogue of unitary operations on the quantum part and 
detection of light by the detector is the analogue of the measurement operation.
Wave amplitude serves as a parallel to the complex probability amplitudes and 
wave phase serves as a parallel to the relative phase among the probability amplitudes.

This paper is organized as follows. The next section gives a brief introduction to some principles of optics pertinent to our model. 
In section 3 we define the model. 
Section 4 presents interference automata for recognizing the following languages: 
\begin{enumerate}
\item $L_{centre}= \{w_1aw_2\mid w_1,w_2 \in \{a,b\}^*,|w_1|=|w_2|\}$. 
\item $L_{eq}=\{a^nb^n \mid n \in \mathbb{N}\}$. 
\item $L_{pal}=\{ww^R\mid w\in \{a,b\}^*$ where $w^R$ is $w$ reversed$\}$.
\item $L_{bal}=\{w\in \{(,)\}^* \mid $ parentheses in $w$ are balanced$\}$.
\item $L_{sq}=\{a^nb^{n^2} \mid n \in \mathbb{N}\}$. 
\item $L_{pow}=\{a^nb^{2^n} \mid n \in \mathbb{N}\}$. 
\end{enumerate}
Section 5 shows the existence of a language that no interference automata can recognize, but 
that can be recognized by an $O(n^3)$ space Turing machine.
The final section closes with a discussion and some open problems.

\section{A Brief Introduction to the Physics of Light}\label{intersec2}

We now briefly discuss the mathematical 
formalism for interference of monochromatic light of wavelength $\lambda$.
For excellent expositions on the phenomenon of optical interference see \cite{Born} and \cite{Feynman}.

The equation of a light wave at a point \begin{bf}p\end{bf} in space can be described as $R= Ae^{i(\omega t + \phi)}$, where $A$ is the amplitude, 
$\omega$ the angular frequency, and $\phi$ the phase associated  with the wave at that point.
The amplitude $A$ at \begin{bf}p\end{bf} is  $A_0/r$, where $r$ is its distance from the source 
and $A_0$ the initial amplitude at $r=0$. 
The \emph{intensity} of the wave is 
the average energy arriving per unit time per unit area. At any given point, it is proportional to the square of the amplitude
of the wave. 
If two (or more) waves exist at the same point in space, they \emph{interfere} with each other and give rise to a new resultant 
wave. 

Suppose the waves have the same angular frequency and are described by $R_1= A_1e^{i(\omega t + \phi_1)}$ and
$R_2= A_2e^{i(\omega t + \phi_2)}$. 
Then, the resultant is given by $R=R_1+R_2= A_1e^{i\omega t} e^{i\phi_1} +  A_2e^{i\omega t}e^{ \phi_2}$.
The amplitude of the resultant wave is the length of $R$. Since $R$ is a complex number, the amplitude is $\sqrt{R\bar{R}}$. 
Thus, the intensity associated with the wave is proportional to $A_1^2 + A_2^2 + 2A_1A_2\cos(\phi_2 -\phi_1)$.

If two interfering waves have equal amplitude and a phase difference  $\phi_2 -\phi_1=\pi$, 
the resultant intensity at that point is zero. A light detector placed at that point fails to detect any light. 
To use this formalism for calculating the intensity at a point, only the difference $\phi_2 -\phi_1$ is necessary.
This phase difference might arise because of (a) the intrinsic phase difference at their source, or (b) the difference $\Delta$ 
between the distances traveled by each before reaching that point.
Hence, $\phi_2 -\phi_1 = \alpha + \frac{2\pi}{\lambda}\Delta$ where the first term  is the difference in the intrinsic phases
and the second is the difference between the distances in terms of phase. 

In general, if $n$ monochromatic waves $\{R_i=A_ie^{i\omega t +\phi_i}\}_{i=1}^n$ 
interfere at a point, the resultant wave is described by the vector sum
$R= \sum_{i=1}^nA_ie^{i\omega t +\phi_i}$.
A useful point to note is that if the sources are simply different points
on the same wavefront, they have zero intrinsic phase difference. Moreover, the 
phase of a wave can be shifted as required.

\section{The Model }\label{intersec3}
Informally, the interference automaton proposed in this work consists of a finite control, an optical arrangement and a read-only tape
which contains the input $w$ demarcated by end-markers \textcent~and $\$$ in $|w|+2$ cells.
The optical arrangement consists of a linear array of  $|w|+2$ monochromatic 
light sources, each source corresponding to a tape cell, capable of emitting
light with an initial relative phase of $0$ or $\pi$. 
These sources can only be \emph{toggled}.
The distance between any two consecutive sources is the same, a constant independent of the
size of the input.

We have a detector whose movement is dictated by the finite control. Since the control is finite, the movement 
of the detector has to be discretized. The easiest way to do this is to  imagine that the detector moves
along the lines of a grid placed before the array of sources as shown in figure 3.1.
\begin{figure}
\begin{center}
\includegraphics[width=0.5\textwidth]{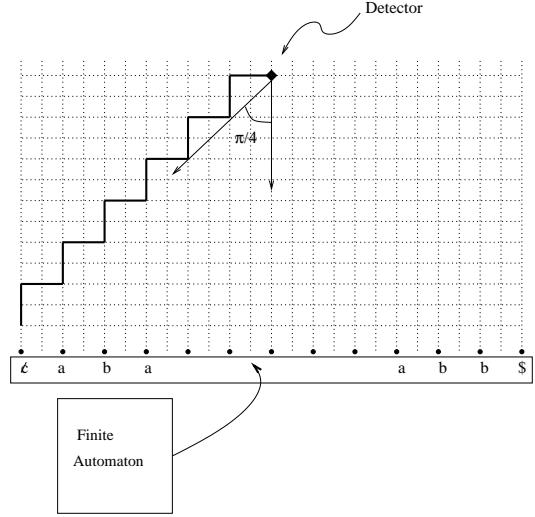}
\caption{The arrangement. Also illustrates the algorithm for recognizing $L_{centre}$. The source at $\$$ comes into the field of vision at the
position where the detector is shown.}
\end{center}
\end{figure}

The detector ``points" in the direction of the source array, parallel to the vertical grid-lines.  
It has a field of vision, which makes an angle $\theta$ with the
vertical gridline passing through it.
We use $\theta = \pi/4$. We associate a coordinate system with the grid by defining horizontal and vertical lines at every half integer point on the horizontal
and vertical axes.
Therefore the vertical gridlines are at $x=0,1/2,1,3/2,\ldots,n+1$ and the horizontal grid-lines are at
$y=0,1/2,1,3/2,\ldots,n+1$. The light sources are placed at $(0,0), (1,0),\ldots, (n+1,0)$. While the sources at $(0,0)$ and $(n+1,0)$
correspond to the end markers~\textcent~and $\$$ respectively, those at  $(1,0),\ldots, (n,0)$ correspond to $w_1,\ldots,w_n$ respectively,
where $w=w_1w_2\ldots w_n$ is the input.

Thus, the position of the tape head is referred to by its x-coordinate. 
The location of the detector is given by $(i,j)$ with $i,j\in \{0,1/2,1,3/2,\ldots,n+1\}$.
We now define the model formally.
\begin{definition}
A 2-way Optical Interference Automaton (2OIA) is defined by a 7-tuple $(Q,\Sigma,Q_{acc},Q_{rej},q_0,\Sigma_I,\delta)$ where
\begin{itemize}
\item $Q$ is a finite set of states.
\item $\Sigma$ is a finite alphabet. 
\item $\Gamma= \Sigma \bigcup \{$ \textcent, $\$\}$ is the tape alphabet.
\item $Q_{acc}, Q_{rej} \subseteq Q$ are sets of accepting states and rejecting states respectively.
\item $q_0$ is a special state designated as the start state.
\item $\Sigma_I=\{\underline{0},\underline{1}\}$ is the output alphabet of the detector.
\item 
The \emph{Instantaneous Description} of a 2OIA is given by the tuple
$(q,i,(j,k),S)$, where $q\in Q$, $i\in \{0,\ldots n+1\}$ is the position 
of the head on the tape, $(j,k)$, $j,k \in \{0,1/2,1,3/2,\ldots,n+1\} $ gives the
position of the detector, and $S$ is a vector of length $n+2$ with entries
from $\{0,\pi,\times\}$, signifying if a source corresponding to a
tape cell is switched on with phase $0$ or $\pi$, or switched off, respectively.

$\delta$ is a transition function--

$\delta :Q\times \Gamma \times \Sigma_I  \rightarrow Q  \times D_t \times D_d \times (toggle(\phi)\cup -)$ 

where $D_t=\{\textrm{left, right, stationary}\}$, $D_d =\{\textrm{left, right, up, down, stationary}\}$ and 
\begin{displaymath}
toggle(\phi) = \left\{ \begin{array}{ll}
\textrm{ switch on the current source with phase $\phi$ } & \textrm{ if the source is off}\\
\textrm{ switch the current source off} & \textrm{ if the source is already on}
\end{array} \right.
\end{displaymath}
where $\phi \in \{0,\pi\}$ and ``--" indicates that no action is to be taken on the current source.
By current source we mean the source associated
with the tape cell currently being scanned by the head.

If $\delta(q,\sigma,\sigma_I) = (q',d_h,(d_l,d_u),p)$
then we denote the change in the instantaneous description during this \emph{one} step as 
$(q,i,(j,k),S)\vdash (q',i',(j',k'),S')$ where $i'=i+d_h$, $j'=j+d_l$, $k'=k+d_u$ 
and 
\begin{eqnarray}
S'[m]&=&S[m] \textrm{ for $m\neq i$ } \nonumber\\ 
     &=&toggle(p) \textrm{ for $m=i$, if $p\in\{0,\pi\}$ }\nonumber\\\nonumber
     &=&S[m] \textrm{ for $m=i$ if $p=-$.}\nonumber\\\nonumber
\end{eqnarray}

For any given source, a \emph{maximal sequence} of toggles is a sequence
over successive time steps $(t_i,t_{i+1}, \ldots,t_j)$ such that either 
(a) the source is not toggled at $t_{i-1}$ and $t_{j+1}$ or,
(b) $t_i=t_0$ or $t_j=t_T$ where $t_0$ and $t_T$ are the time steps at which
computation starts and ends, respectively.
A maximal toggle sequence is called \emph{transient} if its length is even and 
\emph{non-transient} if its length is odd.
We place the restriction that non-transient sequences be allowed at most
a constant number $k$ number of times: on attempting a non-transient 
sequence of toggles a $k+1^{th}$ time, the machine \emph{crashes}.

\end{itemize} 
\end{definition}

The detector has an output alphabet through which it can indicate whether
it has detected any light. It responds with a $\underline{1}$ if it has, and with 
a $\underline{0}$ if it has not.

Depending on the current state $q$, the symbol $\sigma\in \Gamma$ currently being scanned, 
and the output of the detector, $\delta$  changes the state
of the finite control to $q'$ and moves the tape head by $d_t\in D_t$  and the
detector by $d_d\in D_d$. 

The primary use of the sources is to produce optical interference at the grid.
However, unrestricted toggling would enable their use as memory elements: the 
detector when close to the array, could ``read" an individual source
while the finite control could ``write" to it by toggling it. In order to 
avoid this, we place restrictions on moves that involve toggle operations.
A source that is toggled twice in quick succession cannot be used as a memory
element, as it is restored to its original (switched on or off) state in the very next time step. 
Therefore, transient toggling of the kind
$(q,i,(j,k),S)\vdash (q',i,(j',k'),S')\vdash(q',i',(j'',k''),S)$
are permitted any number of times.

Non-transient toggles which allow changing the state of a source
for an extended period of time can potentially be used as memory operations.
Hence, we restrict number such sequences to at most a constant for any source.
In all interference automata that we will discuss henceforth, we 
use at most two such non-transient sequences per source during the entire
course of computation. Figure 3.2 shows the example profiles of allowed and 
disallowed sequences of toggle operations on a given source.

\begin{figure}
\begin{center}
\includegraphics[width=0.5\textwidth]{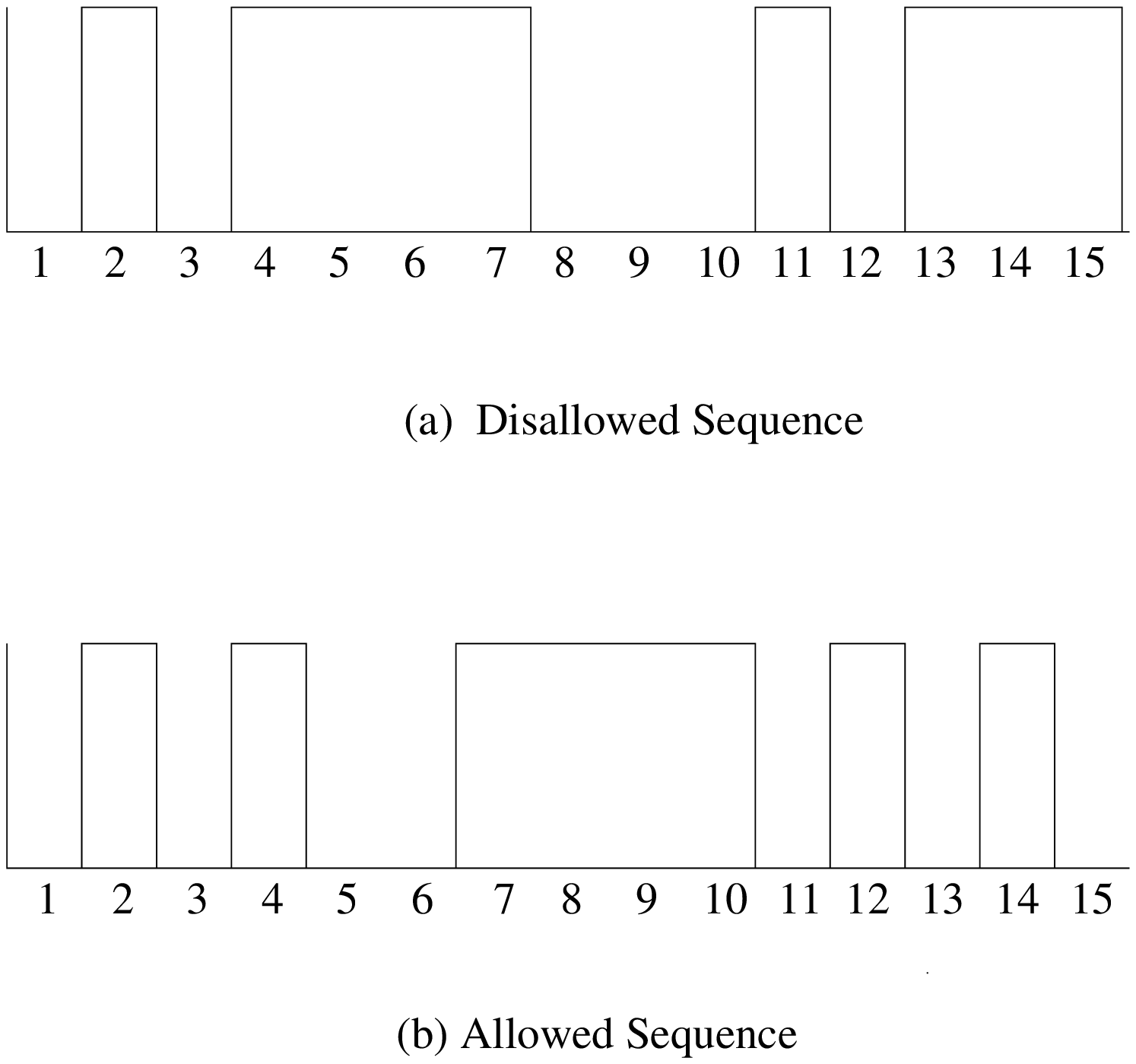}
\caption{Example toggle profiles of a source during the course of computation. 
}
\end{center}
\end{figure}

The sum of the number of moves made by the head and the detector serves as a 
measure of the time taken by a 2OIA machine.

The following facts and conventions will be common to all 2OIA machines that
follow.
\begin{itemize}
\item Initially, all the sources are switched off.
\item $Q_{acc}=\{q_{acc}\}$  and $Q_{rej}=\{q_{rej}\}$ will be the accepting
and rejecting subsets. The machines accept by final state and $\underline{0}$
detector output.
\item The transition function is specified as a table, the rows of which
are indexed by $q\in Q$ and columns by $(\sigma,I)$, with $\sigma \in \Gamma$ and
$I\in \Sigma_I$. The table entries are tuples of the form $(q',t;d,\phi)$ with
$q'\in Q$, $t\in D_t=\{L,-,R\}$, $d\in \{\leftarrow,\rightarrow,\uparrow,\downarrow,\nearrow,\nwarrow,\searrow,\swarrow,-\}$ and $\phi \in\{0,-,\pi\}$.
In any field of a tuple, ``$-$" stands for  a ``no change" in its value. 
Although $D_t$ and $D_d$ imply movement by only one unit (tape cell and grid-line respectively),
movement by two units can be carried out by the finite control. 

An element indexed by row $q$ and column $(a,\underline{1})$, 
say, $(q',R;\nearrow,0)$ is to be interpreted as 
follows. If the machine is in state $q$ and the 
head is reading an $a$, and the detector outputs $\underline{1}$,
then the source corresponding to the current cell is switched on with
a phase $0$, tape head moves right to the next tape cell, the detector moves 
first upwards and then to the right by two grid lines, and the machine enters 
state $q'$. In some places we show both movements in the same entry for the sake 
of brevity, but the two are not atomic. If the detector output changes after
moving up but before moving to the right, the configuration changes as dictated
by the transition function.
\end{itemize}

\section{Language recognition}\label{intersec4}
In this section we show recognition schemes for the languages mentioned in 
Section 1. 
In all cases we show the existence of a locus of points on the grid where the intensity of light will
be zero if and only if the input word is in the language. The idea behind all algorithms on this model is to move
the detector to the locus.

We begin with a simple but important example.

\subsection{Recognizing the centre of an input string}

A word in $L_{centre}= \{w_1aw_2\mid w_1,w_2 \in \{a,b\}^*,|w_1|=|w_2|\}$ will be  of odd length (say $2n+1$). This can be easily verified with a DFA. 
We accept the string, if the $(n+1)^{th}$ element is $a$. 

\begin{theorem}
There exists a 2OIA machine that recognizes $L_{centre}$ in time linear in size of the input string.
\end{theorem}
\begin{proof}
As proof, we describe such a machine. Consider a 2OIA $M_{centre}$ having a set of
states $Q=\{q_0,q_1,q_2, q_{rej},q_{acc}\}$ and a transition function as specified
by the following table.

\begin{tabular}{|c|c|c|c|c|}
\hline
\hline
 &\textcent,\underline{0} & \textcent,\underline{1} & $a$,\underline{0} & $a$,\underline{1} \\
\hline
$q_0$    &  $(q_0,R;-,0)$   &   $\phi$   &  $\phi$ &    $(q_0,R;-,-)$ \\
\hline
$q_1$& $\phi$  & $(q_2,-;-,-)$ & $\phi$ &  $(q_1,L;-,-) $ \\
\hline
$q_2$& $\phi$  & $(q_2,R;\nearrow,-)$ & $(q_{acc},-;-,-)$ & $(q_2,R;\nearrow,-)$  \\
\hline
\end{tabular}

\begin{tabular}{|c|c|c|c|c|}
\hline
& $b$, \underline{0}& $b$,\underline{1} & $\$$, \underline{0} & $\$$, \underline{1} \\
\hline
$q_0$& $\phi$ &   $(q_0,R;-,-)$  &   $\phi$ &     $(q_1,L;-,\pi)$ \\
\hline
$q_1$ & $\phi$ & $(q_1,L;-,-)$&  $\phi$ & $\phi$\\
\hline
$q_2$ & $(q_{rej},-;-,-)$ & $(q_2,R;\nearrow,-)$ & $\phi$ & $\phi$ \\
\hline
\hline
\end{tabular}

The machine starts with the detector initially in position $(0,1)$ and the tape head at $0$, reading~\textcent. 
While in state $q_0$, the sources at~\textcent~and $\$$ are switched on with phase
$0$ and $\pi$ respectively in one rightward scan of the input after which the head
returns to the beginning of the input.

To begin with, the source corresponding to
$\$$ is out of the detector's field of vision. 
However, the intensity at the detector due to the source at~\textcent~is non-zero. 
The detector then moves upwards and to the right 
in such a way that at all times during the movement, it can detect light from the source at~\textcent. 
The head also moves to the right \emph{in tandem} with the detector.
When the detector reaches a certain $x$-coordinate, the source
at $\$$ falls into the detector's field of vision and the resultant intensity falls to zero.
We claim that this coordinate is the geometric centre of the input.
\begin{lemma}
The detector will record zero intensity  if and only if its x-coordinate is $n+1$.
\end{lemma}
\begin{proof}
The resultant wave at the detector is 
\[
\psi = \frac{A_0}{\sqrt{i^2 + j^2}}e^{i\phi_1} + \frac{A_0}{\sqrt{(2n +2 -i)^2 + j^2}}e^{i\phi_2}
\]
for $j\geq 1$. Given that $\phi_2 = \phi_1+\pi$, the detector reads zero resultant intensity if and only if its x-coordinate is $n+1$. $\square$
\end{proof}
Therefore at this moment, if the head reads an $a$, the machine accepts; otherwise it rejects.

It is easy to see that the automaton takes $O(n)$ time to decide the language. 

\end{proof}
Using the ideas behind the 2OIA algorithm for $L_{centre}$, we can recognize a related language, namely $L_{eq}=\{a^nb^n \mid n \in \mathbb{N}\}$.
This language can be recognized with bounded error on 2QCFA \cite{AW}.
\begin{corollary}
The above algorithm can be used to recognize $L_{eq}$ in linear time.
\end{corollary}
\begin{proof} The DFA first verifies that the input is indeed of the form $a^*b^*$
and that the input is indeed of even length. Next, the centre is detected 
using the above algorithm. The input is of even length and therefore, the detector has to be initially positioned at $(1/2,1)$.
The DFA then checks if the current symbol being read is $a$ and that to the right is $b$.
If it is not, reject the input, else accept. 
This too takes time linear in the input size.
\end{proof}

\subsection{Palindromes}
In this section we give a 2OIA machine that recognizes the language $L_{pal}=\{ww^R\mid w\in \{a,b\}^*$ where $w^R$ stands for $w$ reversed$\}$. 
We use light of wavelength $r\pi$ where $r$ is any algebraic number greater than 
zero.
This restriction on the wavelength yields algorithms that are more elegant and faster, and brings out the main features of the model better.
Later in the section we give a machine that recognizes $L_{pal}$ without
this restriction.

\begin{theorem} 
There exists a 2OIA machine that decides $L_{pal}$ in time linear in the input size.
\end{theorem}
\begin{proof}
We describe a 2OIA $M_{pal}$ that decides $L_{pal}$. The automaton $M_{pal}$ has the state
set $\{q_0\ldots q_{10}, q'_3\ldots q'_{10},q_{acc},q_{rej}\}$ and its transition function is as shown in
the following table.

\begin{longtable}{|c|c|c|c|c|}
\hline
\hline
 &\textcent,\underline{0} & \textcent,\underline{1} & $a$,\underline{0} & $a$,\underline{1} \\
\hline
$q_0$    &  $(q_0,R;-,0)$   &   $\phi$   &  $\phi$ &    $(q_0,R;-,-)$ \\
\hline
$q_1$& $\phi$  & $(q_2,-;-,-)$ & $\phi$ &  $(q_1,L;-,-) $ \\
\hline
$q_2$& $\phi$  & $(q_2,R;\nearrow,-)$ & $(q_3,R;-,0)$ & $(q_2,R;\nearrow,-)$  \\
\hline
$q_3$  & $\phi$ & $\phi$ & $\phi$ & $(q_4,L;-,-)$\\
\hline
$q_4$  & $\phi$ & $\phi$ & $(q_5,L;\uparrow,-)$ & $(q_4,-;\downarrow,-)$ \\
\hline
$q_5$  &  $\phi$ & $(q_6,R;-,-)$ & $\phi$ &$(q_5,L;-,-)$  \\
\hline
$q_6$  & $\phi$ & $\phi$ & $(q_7,R;-,-)$ & $(q_6,R;-,0)$ \\
\hline
$q_7$  &  $\phi$  &  $\phi$ & $(q_7,R;-,\pi)$  &  $\phi$  \\
\hline
$q_8$  & $\phi$ &  $\phi$  &  $\phi$  &  $\phi$ \\
\hline
$q_9$  & $\phi$ &  $\phi$  &  $\phi$  &  $\phi$ \\
\hline
$q_{10}$  & $\phi$ &  $\phi$  &  $\phi$  &  $\phi$ \\
\hline
$q'_3$  & $\phi$ & $\phi$ & $\phi$ & $(q_{rej},-;-,-)$\\
\hline
$q'_4$  & $\phi$ & $\phi$ & $\phi$ & $\phi$\\
\hline
$q'_5$  & $\phi$ & $(q'_6,R;-,-)$  & $\phi$& $(q'_5,L;-,-)$\\
\hline
$q'_6$  & $\phi$ & $\phi$ & $\phi$ &$(q'_6,R;-,-)$ \\
\hline
$q'_7$  &  $\phi$  &  $\phi$ & $(q'_7,R;-,-)$  &  $\phi$ \\
\hline
$q'_8$  & $\phi$ &  $\phi$ & $\phi$  &  $\phi$ \\
\hline
$q'_9$  & $\phi$ &  $\phi$ & $\phi$  &  $\phi$\\
\hline
$q'_{10}$  & $\phi$ &  $\phi$ & $\phi$  &  $\phi$\\
\hline

\end{longtable}

\begin{longtable}{|c|c|c|c|c|}
\hline
& $b$, \underline{0}& $b$,\underline{1} & $\$$, \underline{0} & $\$$, \underline{1} \\
\hline
$q_0$& $\phi$ &   $(q_0,R;-,0)$  &   $\phi$ &     $(q_1,L;-,\pi)$ \\
\hline
$q_1$& $\phi$ & $(q_1,L;-,-)$&  $\phi$ & $\phi$ \\
\hline
$q_2$& $(q'_3,R;-,0)$ & $(q_2,R;\nearrow,-)$ & $\phi$ & $\phi$\\
\hline
$q_3$ & $\phi$ &  $(q_{rej},-;-,-)$ & $\phi$ & $\phi$ \\
\hline
$q_4$& $\phi$ & $\phi$& $\phi$& $\phi$ \\
\hline
$q_5$&  $\phi$ & $(q_5,L;-,-)$ &$\phi$& $\phi$\\
\hline
$q_6$&  $\phi$ &$(q_6,R;-,-)$  & $\phi$ & $\phi$ \\
\hline
$q_7$&  $(q_7,R;-,-)$  &  $\phi$ & $(q_8,-;-,-)$  &  $\phi$ \\
\hline
$q_8$&  $\phi$ &  $\phi$ & $(q_9,-;\uparrow,\pi)$ & $\phi$  \\
\hline
$q_9$&  $\phi$ &  $\phi$  & $(q_8,-;-,\pi)$ & $(q_{10},-;-,\pi)$ \\
\hline
$q_{10}$&  $\phi$ &  $\phi$  & $(q_{acc},-;-,-)$ & $(q_{rej},-;-,-)$ \\
\hline
$q'_3$ & $\phi$ & $(q'_4,L;-,-)$ & $\phi$ & $\phi$ \\
\hline
$q'_4$& $(q'_5,L;\uparrow,-)$ & $(q'_4,-;\downarrow,-)$ & $\phi$& $\phi$ \\
\hline
$q'_5$ & $\phi$ &  $(q'_5,L;-,-)$& $\phi$& $\phi$\\
\hline
$q'_6$ & $(q'_7,R;-,-)$ &$(q'_6,R;-,0)$&  $\phi$  & $\phi$\\
\hline
$q'_7$& $(q'_7,R;-,\pi)$  &  $\phi$  &   $(q'_8,-;-,-)$  &  $\phi$\\
\hline
$q'_8$&  $\phi$  &  $\phi$& $(q'_9,-;\uparrow,\pi)$ & $\phi$  \\
\hline
$q'_9$ &$\phi$ & $\phi$  & $(q'_8,-;-,\pi)$ & $(q'_{10},-;-,\pi)$ \\
\hline
$q'_{10}$ &$\phi$ & $\phi$  & $(q_{acc},-;-,\pi)$ & $(q_{rej},-;-,-)$ \\
 \hline
 \hline
\end{longtable}

The automaton, in states $q_0,q_1$ and $q_2$, finds the centre as in the previous section. 
Since the input is of even length, say $2n$, at this stage the head reads the 
$n^{th}$ input symbol $w_n=\sigma \in\{a,b\}$ and the detector is on the 
grid-line $x=n+1/2$. The case of $w_n$ being $a$ is handled by states $q_3\ldots q_{10}$,
while $w_n$ being $b$ is handled by the states $q'_3\ldots q'_{10}$. 
By the arrangement of the sources and the fact 
that $\Sigma =\{a,b\}$, it is sufficient to show symmetry of 
any one $\sigma\in\Sigma$ about the centre of the input string. We show the proof 
for the case when $w_n=a$. The argument for the other case is symmetric.

If $w_n\neq w_{n+1}$, the input is rejected at the outset (see rows indexed by $q_2$ and $q_3$). Else, switch on the source corresponding to $w_n$ and bring the detector to $(n+1/2,1/2)$ so that only
the sources corresponding to $w_n$ and $w_{n+1}$ lie in the field of vision of the detector 
(row $q_4$). This is done by using the fact that as the detector moves down
along the grid-line $x=n+1/2$, the detector falls to $\underline{0}$ when the source
corresponding to $w_n$ moves out of its field of vision, that is, at $y=1/2$.

Then, the head starts from $w_1$, switching
on all sources corresponding to $a$ and with phase $0$ until it reaches $w_n$
(rows $q_5$ and $q_6$). We can deduce when the head has reached the centre because
when the source at $w_{n+1}$ is toggled, the detector records zero intensity. 

As the head moves further to the right, all sources from $w_{n+1}$ to $w_{2n}$ that correspond to $a$ are switched on with relative phase 
$\pi$ (row $q_7$).

Now that all the sources corresponding to $a$ have been switched on appropriately, we move the detector away from the x-axis along the grid-line
$x=n+1$, bringing two sources (one from each side of the centre) into its field of vision at each step. 

Let us now note a useful lemma.

\begin{lemma}
Suppose sources to the left of $x=n+1/2$ are switched on with phase $0$ and those to the right with phase $\pi$. 
Then, the resultant intensity at $x=n+1/2$ is zero 
if and only if either (1) for every $j^{th}$ source to the left of $x=n+1/2$ switched on, the $j^{th}$ source to the right of 
$x=n+1/2$ is also switched on and vice versa or (2) both sources are off.
\end{lemma}

Therefore, as the detector is moved away, if the intensity is non-zero at any step $i$,  we know that the input is not a palindrome, and reject it.
Otherwise, we stop when~\textcent~and $\$$ fall into the field of vision of the detector 
and accept. We can detect when this happens 
by periodically toggling the source corresponding to \$. 
This is done in states $q_8$ to $q_{10}$.
All we need now is a proof of the above lemma.

\begin{proof} (of lemma)
 The $j^{th}$ source from the centre on either side may or may not be switched on, depending on the input.
 Let the boolean variables $a_j$ and $b_j$ indicate whether the $j^{th}$ source is switched on to the left and right of $x=n+1/2$ respectively.
 Thus, the resultant wave at $x=n+1/2$ is
 \[
 (a_1 -b_1)r_1e^{i\phi_1} + (a_2 -b_2)r_2e^{i\phi_2} + \ldots + (a_n -b_n)r_ne^{i\phi_n} 
 \]
where $\phi_j = j\sqrt{2}\pi d/\lambda$ for $1\leq j\leq n$, and $d$ distance between two consecutive vertical (or horizontal) grid-lines. 

Therefore, we have to prove that the resultant is zero if and only if $a_j=b_j$, for $1\leq j\leq n$. One direction is trivial. For the other direction, we use
the following theorem of Lindemann (see~\cite{Niven}).
\begin{theorem}
Given any distinct algebraic numbers $\phi_1, \phi_2,\ldots,\phi_n$, the values $e^{\phi_1}, e^{\phi_2}, \ldots,e^{\phi_n}$ are linearly
independent over the field of algebraic numbers.
\end{theorem}

Since we have chosen $\lambda = r\pi$ for some algebraic number $r$, we have  $\phi_j = j\sqrt{2} d/r$  for $1\leq j\leq n$. 
Therefore, the resultant can be zero if and only if $a_j=b_j$, $1\leq j\leq n$.
\end{proof}

The running time of the algorithm is $O(n)$ as the head scans the input only a constant
number of times and the detector movement is also $O(n)$. 
\end{proof}

Interestingly, $L_{pal}$ can be recognized by a 2OIA without restriction on the
wavelength. However, this comes at the cost of increased time complexity.

\begin{theorem}
There exists a 2OIA machine that recognizes $L_{pal}$ in $O(n^2)$ time.
\end{theorem}
\begin{proof}
Without loss of generality let 
$w_n$ and $w_{n+1}$ be $a$. If the centre is $x=x_c$, then for every $a$ to the left of $x=x_c$ 
we try to find the corresponding $a$ at the same distance to the right, and 
vice versa.
For the sake of brevity, we describe a 2OIA machine $M'_{pal}$ that checks only 
one way:
if for every $a$ to the left of the centre, there exists an $a$ to the right.
In doing so, sources corresponding to $a$'s of only the first half
of the input string will be toggled twice in a non-transient manner.
Therefore, when the extended 2OIA for $L_{pal}$ checks for the existence
of an $a$ to the left of $x_c$ corresponding to every $a$ to the right, the 
sources in the second half of the input can be toggled non-transiently.
The set of states and the transition matrix for the complete case will 
involve a simple and symmetric extension of $M'_{pal}$.

Consider the 2OIA $M'_{pal}$ with states $\{q_0\ldots q_7,q'_3\ldots q'_7,q_{acc},q_{rej}\}$
and transition function as shown in the following table. The detector is
initially at $(1/2,1/2)$.

\begin{longtable}{|c|c|c|c|c|}
\hline
\hline
 &\textcent,\underline{0} & \textcent,\underline{1} & $a$,\underline{0} & $a$,\underline{1} \\
\hline
$q_0$    &  $(q_0,R;-,0)$   &   $\phi$   &  $\phi$ &    $(q_0,R;-,-)$\\
\hline
$q_1$& $\phi$  & $(q_2,-;-,-)$ & $\phi$ &  $(q_1,L;-,-) $\\
\hline
$q_2$& $\phi$  & $(q_2,R;\nearrow,-)$ & $(q_3,-;-,-)$ & $(q_2,R;\nearrow,-)$\\
\hline
$q_3$ & $(q_{acc},-;-,-)$ & $\phi$ & $(q_4,R;-,0)$ &$\phi$ \\
\hline
$q_4$ & $\phi$ & $\phi$& $\phi$ &$(q_5,-;-,\pi)$ \\
\hline 
$q_5$ & $\phi$ & $\phi$ & $(q_6,L;-,\pi)$ & $(q_4,R;-,\pi)$\\

\hline
$q_6$ & $\phi$& $\phi$& $\phi$ & $(q_7,-;-,0)$ \\

\hline
$q_7$ & $\phi$ & $\phi$ & $(q_3,L;-,-)$ & $(q_6,L;-,0)$\\
\hline
$q'_3$ & $(q_{acc},-;-,-)$ & $\phi$ & $ (q'_3,L;-,-)$ & $\phi$\\
\hline
$q'_4$ & $\phi$ & $\phi$& $\phi$
	& $(q'_4,R;-,-)$\\
\hline 
$q'_5$ & $\phi$ & $\phi$ &
$\phi $ & $\phi$ \\

\hline
$q'_6$ & $\phi$& $\phi$& $\phi$& $(q'_6,L;-,-)$\\

\hline
$q'_7$ & $\phi$ & $\phi$ &
$\phi $ & $\phi$\\
\hline
\end{longtable}

\begin{longtable}{|c|c|c|c|c|}
\hline
& $b$, \underline{0}& $b$,\underline{1} & $\$$, \underline{0} & $\$$, \underline{1} \\
\hline
$q_0$& $\phi$ &   $(q_0,R;-,0)$  &   $\phi$ &     $(q_1,L;-,\pi)$ \\
\hline
$q_1$ & $\phi$ & $(q_1,L;-,-)$&  $\phi$ & $\phi$\\
\hline
$q_2$ & $(q'_3,-;-,-)$ & $(q_2,R;\nearrow,-)$ & $\phi$ & $\phi$ \\
\hline
$q_3$&  $ (q_3,L;-,-)$ & $\phi$ & $\phi$ & $\phi$ \\
\hline
$q_4$& $\phi$& $(q_4,R;-,-)$ & $\phi$ & $(q_{rej},-;-,-)$\\
\hline
$q_5$ &$\phi $ & $\phi$  & $\phi$ & $\phi$\\
 \hline
$q_6$& $\phi$& $(q_6,L;-,-)$& $\phi$& $\phi$\\
\hline
$q_7$ & $\phi $ & $\phi$  & $\phi$ & $\phi$ \\
\hline
$q'_3$ & $(q'_4,R;-,0)$ &$\phi$ & $\phi$ & $\phi$ \\
\hline
$q'_4$ & $\phi$ &$(q'_5,-;-,\pi)$ & $\phi$ & $(q_{rej},-;-,-)$\\ 
\hline
$q'_5$& $(q'_6,L;-,\pi)$ & $(q'_4,R;-,\pi)$ & $\phi$ & $\phi$
\\
\hline
$q'_6$& $\phi$ & $(q'_7,-;-,0)$ & $\phi$& $\phi$\\
\hline
$q'_7$ & $(q'_3,L;-,-)$ & $(q'_6,L;-,0)$& $\phi$ & $\phi$\\
 \hline
 \hline
\end{longtable}

The automaton begins by finding the centre of the 
input (rows $q_0,q_1$ and $q_2$).

Note that at this point, the detector is at $(n+1/2, n+1/2)$.
By lemma 1, if only two sources are switched on, and with opposite phases, 
the detector reads $\underline{0}$ if and only if they are equidistant from the centre.
Thus, searching for matching pair involves the following steps:
\begin{itemize}
\item For an $a$ to the left, switched on with $0$, search right for the 
corresponding $a$ as follows. On encountering an $a$, the corresponding source is 
switched on (the source being off initially, toggling switches it on) 
with a phase $\pi$.
If the detector reads $\underline{0}$, we have found the $a$ that we were
looking for, and the head returns left.
Otherwise, we continue to search to the right. In any case, the source 
is switched off again.
If the head hits \$ without the detector reading a $\underline{0}$ it implies
that input is not a palindrome and is rejected (rows $q_3,q_4$ and $q_5$ for $w_n =a$ and $q'_3,q'_4$ and $q'_5$ for $w_n=b$).
\item 
If the $a$ that we were looking for in the previous step is found,
the corresponding source is switched off (making the detector again read $\underline{1}$),
and the head returns to the left $a$. The catch here is that the finite
control cannot remember the position of the left $a$.
However, since the source 
corresponding to the right $a$ has been switched off, the only source that is 
switched on is the one corresponding to the left $a$. Thus, during the leftward 
scan, if toggling a source corresponding to an $a$ drops the detector reading 
to $\underline{0}$ again, we have found the left $a$ (rows $q_6$ and $q_7$ for $w_n =a$ and $q'_6$ and $q'_7$ for $w_n=b$).
\item Search left for another $a$. If found, repeat the above process. If not,
that is, if the head hits~\textcent, accept (row $q_3$ for $w_n=a$ and $q'_3$ for
$w_n=b$). 
\end{itemize}

Since every symbol in the input is scanned at most $O(n)$ times, and the detector
moves by only $O(n)$ steps during the execution, the machine takes $O(n^2)$ time.

\end{proof}

\subsection{Balanced Parentheses}
We now show a 2OIA machine that recognizes $L_{bal}$ using a combination of 
the techniques for palindromes given in the previous subsection.

\begin{theorem}
There exists a 2OIA machine that recognizes $L_{bal}$ in $O(n^3)$ time, where $n$ is the 
input length.
\end{theorem}
\begin{proof}
Let us first define two useful terms.

\begin{definition}
A string in $\{(,)\}^*$ is called a \emph{simple nest} if it consists of $n$ $`($'s followed by $n$ $`)$'s, for $n\geq 1$.
\end{definition}

\begin{definition}
A string in $\{(,)\}^*$ is called a \emph{compound nest} if it consists of $n$ 
$`($'s followed by  $m$ simple or compound nests, followed by $n$ $`)$'s, 
for $m>1$ and $n\geq 1$.
\end{definition}

A string in $L_{bal}$ is a concatenation of these two types of substrings.
We define a 2OIA $M_{bal}$ that deals with the two cases separately.
It has set of states $Q=\{q_0\ldots q_{15},q_{rej},q_{acc}\}$ and transition function as shown in the following table.

\begin{longtable}{|c|c|c|c|c|}
\hline
\hline
 &\textcent,\underline{0} & \textcent,\underline{1} & $($,\underline{0} & $($,\underline{1} \\
\hline
$q_0$ & $(q_0,R;-,0)$ & $\phi$ & $\phi$& $(q_0,R;-,0)$ \\
\hline
$q_1$ & $\phi$ & $(q_2,R;-,-)$& $\phi$ & $(q_1,L;-,-)$\\
\hline
$q_2$ &$\phi$ &$\phi$ &$\phi$& $(q_2,R;\rightarrow,-)$\\
\hline
$q_3$ & $(q_{rej},-;-,-)$ &$(q_4,R;\downarrow,-)$ & $(q_3,L;\uparrow,-)$& $(q_4,R;\downarrow,-)$\\
\hline
$q_4$ &  $\phi$ & $\phi$  & $(q_4,R;-,0)$& $(q_4,R;-,0)$ \\
\hline
$q_5$ & $\phi$ & $\phi$& $(q_{11},-;-,-)$  & $\phi$\\
\hline
$q_6$& $\phi$&$\phi$&$\phi$&$\phi$\\
\hline
$q_7$ &$(q_{rej},-;-,-)$ &$\phi$&$\phi$ & $(q_{8},-;-,0)$\\
\hline
$q_{8}$ &$\phi$&$\phi$ &$(q_{8},R;-,-)$&$(q_7,L;-,0)$\\
\hline
$q_{9}$ &$\phi$&$\phi$ &$(q_{10},-;-,0)$&$\phi$\\
\hline
$q_{10}$ &$\phi$&$\phi$& $(q_{11},L;\rightarrow,0)$ &$(q_{9},R;-,0)$\\
\hline
$q_{11}$ &$\phi$&$\phi$&$\phi$&$\phi$\\
\hline
$q_{12}$ &$\phi$&$\phi$&$\phi$&$\phi$\\
\hline
$q_{13}$ &$(q_{acc},-;-,-)$&$\phi$& $(q_{13},L;-,-)$ & $(q_{rej},-;-,-)$\\
\hline

\end{longtable}

\begin{longtable}{|c|c|c|c|c|}
\hline
& $)$, \underline{0}& $)$,\underline{1} & $\$$, \underline{0} & $\$$, \underline{1}\\
\hline
$q_0$& $\phi$ &$(q_0,R;-,\pi)$ & $\phi$& $(q_1,\leftarrow, (-,-),-)$  \\
\hline
$q_1$ & $\phi$& $(q_1,L;-,-)$& $\phi$& $\phi$\\
\hline
$q_2$&  $(q_3,L;-,-)$& $(q_2,R;\rightarrow,-)$& $\phi$ & $(q_{13},L;-,-)$\\
\hline 
$q_3$& $(q_4,R;\downarrow,-)$ & $(q_4,R;\downarrow,-)$&$\phi$ &$\phi$ \\
\hline
$q_4$ & $\phi$ & $(q_5,-;-,-)$  & $\phi$  & $\phi$ \\
 \hline
$q_5$ & $(q_6,-;\uparrow,-)$& $(q_5,R;-,\pi)$  & $(q_{13},L;-,-)$ & $\phi$ \\
\hline
$q_6$&$(q_7,-;-,\pi)$ & $(q_6,-;\nwarrow,-),-)$ &$\phi$ &$\phi$\\
\hline
$q_{7}$ &$\phi$ & $(q_7,L;-,-)$ &$\phi$&$\phi$\\
\hline
$q_{8}$ &$(q_{9},R;-,-)$ &$\phi$ &$\phi$&$\phi$ \\
 \hline
$q_{9}$ & $(q_{10},-;-,\pi)$ &$\phi$ & $(q_{13},L;-,-)$ &$\phi$\\
 \hline  
$q_{10}$& $(q_6,-;\uparrow,\pi)$& $(q_{8},-;-,\pi)$&$\phi$&$\phi$\\
\hline
$q_{11}$ &$\phi$& $(q_{12},-;-,\pi)$ &$\phi$&$\phi$ \\
\hline
$q_{12}$& $(q_2,R;-,\pi)$ & $(q_{12},-;\searrow,-)$&$\phi$&$\phi$ \\
\hline
$q_{13}$& $(q_{13},L;-,-)$ &$\phi$&$\phi$&$\phi$\\
\hline
\hline
\end{longtable}

Initially, the sources corresponding to the $`($'s and $`)$'s are switched on with phase $0$ and $\pi$ respectively (while in state $q_0$).
The head reads~\textcent~and the detector is placed at $(1/2,1/2)$.

The head moves to the right ignoring the $`($'s on the way, with the detector moving
in tandem (row $q_2$).

Simple nests are dealt with as follows:
\begin{itemize}
\item Since every $`($' is switched on with phase $0$ and $`)$' with $\pi$, 
if the intensity at the detector is zero when the head is reading a $`)$', 
we conclude that the detector's current x-coordinate is the centre of a
simple nest. 
\item The detector moves up and the head to the left until the leftmost 
$`($' and the rightmost $`)$' of the nest lie in the field of vision of the detector.
Thus, the movement stops when (a) the detector outputs a $\underline{1}$
and/or (b) the head reads~\textcent~or a $`)$'.
If the detector outputs $\underline{0}$ while the head is reading~\textcent,
it implies imbalance: since the source for \$ is not switched on, it means
that~\textcent~has been balanced by a $`)$'. Therefore, the input is rejected.
If the detector outputs a $\underline{1}$, the edge of the current nest has been detected. 
The head moves one cell to the right and the detector one step down, again
outputing a $\underline{0}$ (rows $q_2, q_3$ and $q_4$).
\item Now the head, moving right, toggles (in effect, switches off) all the symbols of this nest. By the same argument as in the previous section, the detector
output changes to $\underline{1}$ as soon as the first symbol is toggled.
It reverts to $\underline{0}$ only when all the symbols corresponding
to this nest are switched off (rows $q_4$ and $q_5$).
\end{itemize}
An important point to note is that the sources corresponding to
a ``recognized nest" (simple or compound) are never switched on in a non-transient
 manner again during computation.
We now turn to compound nests. 
For the rest of the proof we abuse the  notation a bit by calling those 
parentheses in the compound nest that are not a part of an inner simple nest as the compound nest itself. 
A compound nest is recognized as follows.
\begin{itemize}
\item
To begin with, the head reads the innermost $`)$' of the compound nest, say at 
cell $x$, and the detector is at $(x,1/2)$. 
The detector travels to the left and away from the source array until it reaches a position on the grid where it reads $\underline{0}$ (row $q_6$).
This is easy because of the fact that all sources corresponding to simple nests and smaller compound nests contained inside the current
compound nest have been switched off.
\item
Once the centre $x_c$ of the innermost $`($' and  $`)$' of the compound
nest has been located, the head moves to the left to detect the corresponding 
$`($' and switch it off. At this instant, when the innermost pair of the compound
nest has been discovered and switched off, the detector reads a $\underline{0}$.
If the head hits~\textcent~before this happens, it implies imbalance, namely excess $`)$'s and the machine rejects.
The head returns to the right to the $`)$' of the pair (rows $q_7$ and  $q_8$).
These two steps are performed in the same way as described earlier for
 simple nests.

\item If the next symbol is again a $)$, we repeat the above two steps.
If not, the detector has to be brought into  position $(x-1/2,1/2)$, where $x$ 
is the x-coordinate of the next symbol (rows $q_{9}\ldots q_{12}$). If it is $`($', this is the beginning 
of a new nest. If it is \$, we have accounted for all $`)$'s.
All we need to do now is to check if any $`($' is left unpaired.
At this point, all balanced parentheses are switched off. Thus, if during a
leftward scan, the detector outputs $\underline{1}$, we conclude that there exists
an extra $`($' and reject the input. If however, the detector does not output a 
$\underline{1}$ before reaching~\textcent, we accept the input (row $q_{13}$).
\end{itemize}

Therefore, the machine accepts if and only if the input has balanced parentheses. 

Recognizing a simple nest of $2m$ symbols takes at most $O(m)$ moves of the
head and $O(m)$ moves of the detector. 
For a compound nest of $m$ symbols, having $l$ symbols of enclosed simple nests,
at most $O(m^2l)$ moves of the head and $O(l)$ moves of the detector are required.
Since $m$ and $l$ are bounded from above by $n$, the machine takes at most 
$O(n^3)$ time.
\end{proof}

\subsection{$a^nb^{n^2}$}
\begin{theorem} There exists a 2OIA that recognizes $a^nb^{n^2}$ in $O(|w|^3)$ time, where $|w|$ is the length of the input.
\end{theorem}

\begin{proof}
Intuitively, the 2OIA works as follows. It ``measures out" blocks of $b$'s of length $n$
and keeps a count of the blocks measured out thus far using the $a$'s.
For every source corresponding to an $a$ switched on, we measure out and mark a block
of $b$'s. The first and last source of each block is switched on with phase 
$0$ and $\pi$ respectively and the interference at the centre of the block
last measured out is used to measure out the next block. An input $x$ is in the language if and only if the total number of blocks measured
out is exactly equal to the number of $a$'s in $x$. 

The 2OIA consists of states $\{s_0,s_1,s_2\}\cup \{f_1,\ldots,f_7\}\cup
\{b_1,\ldots,b_6\}\cup\{f'_1,\ldots,f'_6\}\cup \{q_{acc},q_{rej}\}$.
The transition function is defined as follows:

\begin{longtable}{|c|c|c|c|c|}
\hline
\hline
 &\textcent,\underline{0} & \textcent,\underline{1} & $a$,\underline{0} & $a$,\underline{1}\\
\hline
$s_0$ & $(s_0,R;\rightarrow,-)$ & $\phi$ & $(s_0,R;-,0)$& $(s_0,R;-,-)$\\
 \hline
 $s_1$ &  $\phi$ & $\phi$& $\phi$ & $(s_2,R;-,\pi)$\\
 \hline
 $s_2$  &  $\phi$ & $\phi$& $\phi$& $\phi$\\
 \hline
 $f_1$ &  $\phi$ & $\phi$& $\phi$& $\phi$\\
 \hline
 $f_2$ &  $\phi$ & $\phi$& $\phi$& $\phi$\\
 \hline
 $f_3$ &  $\phi$ & $\phi$& $\phi$& $\phi$\\
 \hline
 $f_4$ &   $\phi$ & $\phi$& $\phi$& $\phi$\\
 \hline
 $f_5$ &  $\phi$ & $\phi$&  $\phi$ & $\phi$\\
 \hline
 $f_6$ &  $\phi$ & $\phi$&  $\phi$ & $\phi$\\
 \hline
 $f_7$ &  $\phi$ & $\phi$&  $\phi$ & $\phi$\\
 \hline
 $b_1$ &  $\phi$ & $(b_2,R;\rightarrow,-)$ & $(b_1,L;\leftarrow,-)$ & $(b_1,L;\leftarrow,-)$\\
 \hline
 $b_2$ & $\phi$ & $\phi$&  $(b_3,R;\rightarrow,0)$ & $(b_2,R;\rightarrow,-)$\\
 \hline
 $b_3$ &  $\phi$ &$\phi$& $(b_4,R;\rightarrow,-)$  & $(b_6,-;-,-)$\\
 \hline
 $b_4$ &  $\phi$ &$\phi$& $(b_4,R;\rightarrow,-)$  & $(b_4,R;\rightarrow,-)$\\
 \hline
 $b_5$ &  $\phi$ & $\phi$&  $\phi$ & $\phi$\\
 \hline
 $b_6$ &  $\phi$ & $\phi$&  $\phi$ & $\phi$\\
 \hline
 $b_7$ &  $\phi$ & $\phi$&  $\phi$ & $\phi$\\
 \hline
 $f'_1$ &  $\phi$ & $\phi$&  $\phi$ & $\phi$\\
 \hline
 $f'_2$ &  $\phi$&  $\phi$ & $\phi$ & $\phi$\\
 \hline
 $f'_3$ & $\phi$ & $\phi$ &  $\phi$ & $\phi$\\
 \hline
 $f'_4$ &  $\phi$ & $\phi$ &  $\phi$ & $\phi$\\
 \hline
 $f'_5$ & $\phi$ & $\phi$&  $\phi$ & $\phi$\\
 \hline
 $f'_6$ & $\phi$ & $\phi$&  $\phi$ & $\phi$\\
\hline
\hline
\end{longtable}

\begin{longtable}{|c|c|c|c|c|}
\hline
\hline
 &$b$,\underline{0} & $b$,\underline{1} & $\$$,\underline{0} & $\$$,\underline{1}\\
\hline
$s_0$ & $\phi$ & $(s_1,L;-,-)$ & $\phi$ &$(q_{rej},-;-,-)$\\
\hline
$s_1$&$\phi$ &$\phi$ &$\phi$& $\phi$\\
\hline
$s_2$& $(f_2,-;\rightarrow,0)$ & $(s_2,-;\nearrow,-)$ & $\phi$ & $\phi$ \\
 \hline
 $f_1$ &  $(f_2,-;\rightarrow,-)$& $(f_1,-;\nwarrow,-)$ &$\phi$ & $\phi$\\
 \hline
 $f_2$  & $(f_3,-;\rightarrow,-)$& $(f_2,-;\rightarrow,-)$ &$\phi$ & $\phi$\\
 \hline
 $f_3$ &  $(f_4,L;\leftarrow,-)$& $(f_3,R;\rightarrow,-)$ &$\phi$ & $(q_{rej},-;-,-)$\\
 \hline
 $f_4$ &  $\phi$& $(f_5,-;-,\pi)$ &$\phi$ & $q_{rej}$\\
 \hline
 $f_5$ &  $(f_6,R;-,-)$& $(f_4,R;-,\pi)$ &$\phi$ & $\phi$\\
 \hline
 $f_6$ &  $(f_7,L;-,0)$& $\phi$ &$(q_{rej},-;-,-)$ & $\phi$\\
 \hline
 $f_7$ &  $(b_1,-;-,-)$& $(f_7,-;\searrow,-)$ &$\phi$ & $\phi$\\
 \hline
 $b_1$ &  $(b_1,L;\leftarrow,-)$&$(b_1,L;\leftarrow,-)$  &$\phi$ & $\phi$\\
 \hline
 $b_2$ & $\phi$ & $\phi$& $\phi$ & $\phi$\\
 \hline
 $b_3$ &  $\phi$ & $\phi$& $\phi$ & $\phi$\\
 \hline
 $b_4$ & $(b_4,R;\rightarrow,-)$ & $(b_4,R;\rightarrow,-)$ & $(b_5,L;\leftarrow,-)$ &$\phi$\\
 \hline
 $b_5$ & $(b_5,L;\leftarrow,-)$ & $(f_1,-;-,-)$ & $\phi$ &$\phi$\\
 \hline
 $b_6$  & $(b_6,R;\rightarrow,-)$ & $(b_6,R;\rightarrow,-)$ & $(b_7,L;\leftarrow,-)$ &$\phi$\\

 \hline
$b_7$ & $(b_7,L;\leftarrow,-)$ & $(f'_1,-;-,-)$ & $\phi$ &$\phi$\\

 \hline
 $f'_1$ &  $(f'_2,-;\rightarrow,-)$& $(f'_1,-;\nwarrow,-)$ &$\phi$ & $\phi$\\
 \hline
 $f'_2$ &  $(f'_3,-;\rightarrow,-)$& $(f'_2,-;\rightarrow,-)$ &$\phi$ & $\phi$\\
 \hline
 $f'_3$ & $(f'_4,L;\leftarrow,-)$& $(f'_3,R;\rightarrow,-)$ &$\phi$ & $(q_{rej},-;-,-)$\\
 \hline
 $f'_4$ &  $\phi$& $(f'_5,-;-,\pi)$ &$\phi$ & $(q_{rej},-;-,-)$\\
 \hline
 $f'_5$ & $(f'_6,R;-,-)$& $(f'_4,R;-,\pi)$ &$\phi$ & $\phi$\\
 \hline
 $f'_6$ & $q_{rej}$ & $\phi$ & $(q_{acc},-;-,-)$&$\phi$\\
\hline
\hline
\end{longtable}

To begin with, the sources corresponding to the first and last $a$ are switched on with phase $0$
and $\pi$ respectively, and the first block of $b$'s is measured out and marked.
\begin{figure}\label{f2}
\begin{center}
\includegraphics[width=0.9\textwidth]{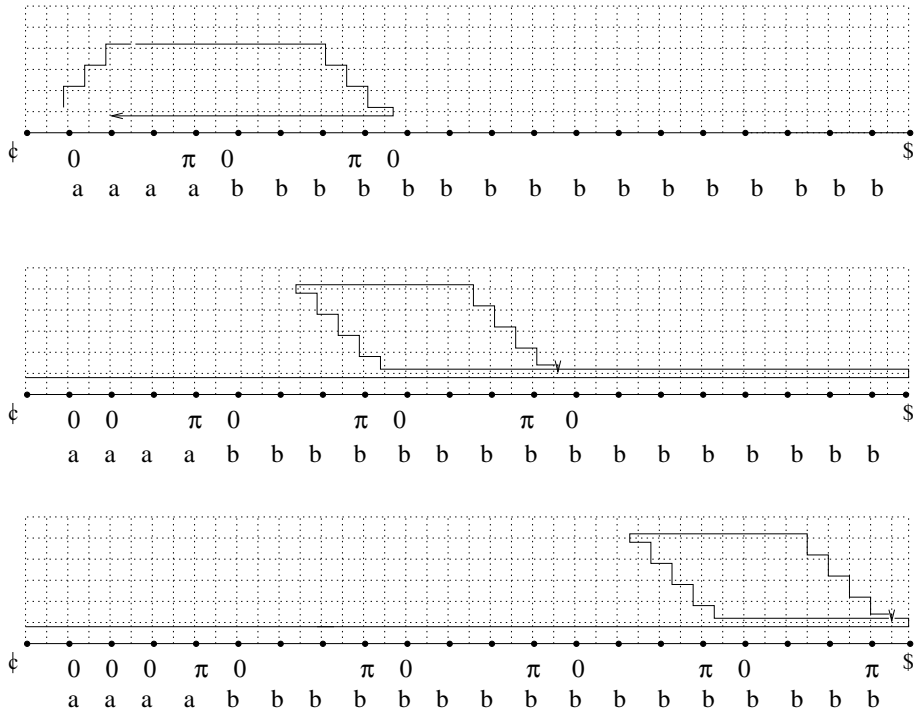}
\caption{Recognizing $a^nb^{n^2}$. The bold lines along the grid-lines show
the trajectory of the detector during various phases.}
\end{center}
\end{figure}

The algorithm consists of several iterations, each of which consists
of the following steps:

\begin{enumerate}
\item {\bf Bookkeeping phase (states $b_1-b_7$):} The $i^{th}$ ($i>1$) iteration begins by switching the source corresponding
to the $i^{th}$ $a$ with phase $0$. The detector is close to the source array
so that it can read each source individually. Starting from the beginning
of the input, the detector and the source are moved to the right in tandem, 
ignoring sources that are already
switched on. The first source that registers a \underline{0} at the detector is 
switched on.
That done, the detector and the head set out to find the rightmost unmarked block,
which actually corresponds to this $a$.
This is accomplished by first travelling to the end of the input string, again
with the detector and head in tandem, and returning left until the head reads
a $\underline{1}$. The coordinates of the 
head and the detector at this moment are $(in+1)$ and $(in+1/2,1/2)$ respectively.
Thus, the $(i-1)^{th}$ block is used for marking out the $i^{th}$ block (see next step).
The first iteration, in which the first block of $b$'s is marked out, uses the 
block of $n$ $a$'s, (states $s_0-s_2$) in a similar fashion.

\item {\bf Marking the next block (states $f_1-f_5$ and $f'_1-f'_6$ for the last block):} The detector is taken to 
$((i-1)n+(n+1)/2,(n+1)/2)$, 
the centre of the $(i-1)^{th}$ block. Now, the right end of the $i^{th}$
block is detected as follows. The detector moves to the right. The first
$\underline{0}$ registered is ignored and the detector keeps moving
to the right till the detector reads a $\underline{0}$ for the second time.
This happens when the source at $in+1$ falls out of the field of vision of the detector.
The detector is moved back by $1/2$ so that the source again falls into the field of vision
of the detector and it registers a $\underline{1}$. The detector is therefore
at the centre of the next block, that is, the block that is to be marked.
The head is now advanced to the right toggling every source on the way
by phase $\pi$. If the detector records $\underline{0}$, we have found
the right end and the head is at $(i+1)n$. 
On the other hand, if the detector records $\underline{1}$, the
source is toggled back again: this is not the right end of the block.
If the head hits \$ without recording $\underline{0}$, the input is rejected.

\item{\bf Bringing the detector back in tandem with the detector (states $f_6-f_7$): }
The detector has to be brought to $((i+1)n,1/2)$, so that the detector and
the head can go back in tandem to the start of the input to begin the next 
iteration. This is done by switching the source corresponding to the next
source by phase $0$. The detector is then moved to the right and closer to the 
source array in steps (see figure 3.3). The detector registers $\underline{1}$
as soon as it moves right from the centre of the $i^{th}$ block. However,
it registers $\underline{0}$ again, when its co-ordinate is $(i+1)n-1/2$. But
since it has also been moving closer to the array, by this time, the detector is
at a distance of $1$ unit from it. 

\end{enumerate}

Let the input be $a^nb^{m}$ for some $n,m\in\mathbb{N}$. Three cases arise:
\begin{enumerate}
\item $m=n^2$: After the last $a$
is switched on, the head and the detector set out to find an unmarked
block of $b$'s. If a complete block of $n$ $b$'s is found and the symbol
immediately after the block is \$, the machine accepts.
\item $m<n^2$:
If $m=l^2<n^2$, then the 2OIA is not in the $f'$ states, indicating that the 
machine is not looking for the last block yet. Therefore, if the head hits \$
immediately after marking a block, the input is rejected. If $m$ is not a
perfect square, then the head hits \$ while it is still marking a block,
and the input is rejected.
\item $m>n^2$: There are residual $b$'s even after marking the block
corresponding to the last $a$, and the input is rejected.
\end{enumerate}

Thus, the machine described above accepts a string if and only if
it is of the form $a^nb^{n^2}$ for $n\in\mathbb{N}$. Further, since the head and
the detector make one scan of the input for each $a$, the total time
taken is $O(|w|^3)$.

\end{proof}

\begin{corollary}
There exists a 2OIA that can recognize $a^nb^{2^n}$ in time $O(2^n)$.
\end{corollary}
\begin{proof}
The 2OIA recognizing $a^nb^{2^n}$ works in a similar fashion. The detector
doubles the length of the blocks in each iteration and 
the $a$'s are used for keep track of the number of blocks of $b$'s.
\end{proof}

\section{A Lower Bound}\label{intersec5}

How powerful is this model? Observe that the head can be in $n$ different
positions and the detector can be in $4n^2$ different positions on the grid.
Moreover, in each of these positions it can be reading either a $\underline{0}$
or a $\underline{1}$. Thus, the total number of distinct configurations possible
for the 2OIA is $|Q|n2^{4n^2}$. 

The space hierarchy theorem of complexity states that
\begin{theorem}
For every space constructible function $f:\mathbb{N} \rightarrow \mathbb{N}$, there exists a language $L$ that is 
decidable in space $O(f(n))$ but not in space $o(f(n))$.
\end{theorem}

This immediately leads to a bound for 2OIA.

\begin{theorem}
Let $f:\{0,\}^n\rightarrow \{0,1\}^m$ be an $\Omega(n^3)$ space constructible function. Then,
no 2OIA can recognize the language $L=\{x\#f(x)\}$.
\end{theorem}
\section{Conclusions and Open Problems}\label{intersec6}

We proposed a model of computing based on optical interference
and showed machines of this model that recognize some non-trivial languages.
Our work leaves the following questions open.
\begin{itemize}

\item Does there exist an elegant characterization of this model?
\item If the number of initial phases is more than just two ($0$ and $\pi$), 
then what is the increase in power?
\item How does this model compare with 2QCFA in terms of language recognition?
Does the set of languages recognized by one model include
that recognized by the other? If 2QCFA is strictly more powerful than 2OIA, 
then our results imply that all the languages posed as open for 2QCFA by 
Ambainis and Watrous~\cite{AW} can be recognized by them.
If the inclusion is the other way, then 2OIA is an upper bound on the power
of 2QCFA. In particular, it would imply that $L_{fn}$ can be recognized by no 2QCFA.
\item If the output alphabet of the detector is expanded, that is, if the 
detector can report different levels of intensity, then what is the increase in power?
\end{itemize}
\bibliographystyle{alpha}
\bibliography{../../thesis/bib}

\end{document}